\documentclass[10pt, conference, compsocconf, letterpaper]{IEEEtran}

\pdfoutput=1

\usepackage{amsmath}

\usepackage{graphicx}
\usepackage[noend]{distribalgo}
\usepackage{algorithm}
\usepackage{amssymb}
\usepackage{color}
\usepackage[draft]{fixme}
\usepackage{multirow}
\usepackage{afterpage}
\usepackage{amsmath}
\usepackage{caption}
\usepackage{soul}

\newcommand{\sizefactor}{0.9}

\newcommand{\SC}{\ensuremath{\mathit{SC}}}

\begin{document}

\title{Parallel Deferred Update Replication}

\author{
    \IEEEauthorblockN{Leandro Pacheco, Daniele Sciascia, Fernando Pedone}
    \IEEEauthorblockA{University of Lugano, Switzerland}}
\maketitle

\begin{abstract}

Deferred update replication (DUR) is an established approach to implementing highly efficient and available storage.
While the throughput of read-only transactions scales linearly with the number of deployed replicas in DUR, the throughput of update transactions experiences limited improvements as replicas are added.
This paper presents Parallel Deferred Update Replication (P-DUR), a variation of classical DUR that scales both read-only and update transactions with the number of cores available in a replica.
In addition to introducing the new approach, we describe its full implementation and compare its performance to classical DUR and to Berkeley DB, a well-known standalone database.

\end{abstract}



\section{Introduction}

Many distributed applications rely on a storage service to persist state. 
Consequently, storage is often the Achilles heel of distributed applications: an application will be at most as efficient and as available as the storage service it relies upon.
This paper considers a class of distributed storage systems based on the deferred update replication approach.
Deferred update replication (DUR) combines high availability and high efficiency, and provides an intuitive interface to applications (i.e., serializable transactions).
Due to these characteristics, DUR lies at the core of many data management protocols (e.g., \cite{KA00a, LKPM+05, PMJP+05, Pedone:2003}).

%

\subsection{Deferred update replication}

%
Deferred update replication distinguishes between update transactions and read-only transactions. 
While an update transaction is executed by a single server, all servers certify the transaction and apply its updates to their local database, should the transaction pass certification.
Certification is necessary to ensure serializability in spite of the fact that update transactions execute without any synchronization across replicas.
Read-only transactions do not need to be certified; a replica can serialize a read-only transaction by carefully synchronizing it locally (e.g., using a multiversion database).

The performance benefits of DUR stem from the fact that the execution of transactions can be shared among replicas, which results in better load distribution among servers than other replication techniques, such as primary-backup and state-machine replication.
With state-machine replication, every update transaction must be executed by all servers~\cite{Sch90}. 
Thus, throughput of update transactions cannot be improved by increasing the number of servers; it is limited by what one replica can execute.
With primary-backup replication~\cite{Sto79}, the primary first executes update transactions and then propagates their updates to the backups, which apply them without re-executing the transactions; the throughput of update transactions is limited by the capacity of the primary, not by the number of replicas.

\subsection{The (lack of) scalability of deferred update replication}


In deferred update replication, read-only transactions scale linearly with the size of the system: the transactions that an ensemble of replicated servers can execute per time unit is directly proportional to the number of servers.
This happens because the execution and termination of read-only transactions are entirely local to a replica.
Unfortunately the same does not hold for update transactions.
Although the execution of an update transaction is local to a server, its termination requires work from each replica (i.e., certifying the transaction and possibly applying its changes to the database).
Therefore, update transactions do not scale with the number of replicas deployed.
Increasing the number of replicas (\emph{scale-out}) can improve update throughput, since more replicas can share the load of transaction execution, but the improvement is limited and not proportional to the added resources.
One can boost update throughput more effectively by upgrading the servers (\emph{scale-up}).
Since modern servers increase processing power by aggregating multiple processors (e.g., multi-core architectures), scaling up update transactions in DUR requires  
supporting parallelism (multithreading) in the execution and termination of transactions.

\subsection{Parallel Deferred Update Replication}

While multithreaded execution of transactions is relatively simple (e.g., using locks for synchronization), multithreaded termination in DUR
is challenging.
Two aspects contribute to the difficulty of designing an efficient multithreaded transaction termination procedure:
First, the procedure must be deterministic since all replicas must reach the same decision after certifying the same transactions.
Second, certification may require synchronization among threads.
We illustrate this last constraint with an example.
Let $t_i$ and $t_j$ be two transactions delivered for certification at a server.
Certification of $t_i$ (resp., $t_j$) essentially checks whether $t_i$ can be serialized with transactions that executed concurrently with $t_i$ but committed before $t_i$'s certification.
Although $t_i$ can be certified against committed transactions by one thread while $t_j$ is certified against committed transactions by another thread, ultimately one transaction must be serialized with respect to the other, which requires synchronizing the certification of $t_i$ and $t_j$. 

In this paper, we present Parallel Deferred Update Replication (P-DUR), an extension to classical deferred update replication that under certain workloads allows update transactions to scale linearly with the number of cores available in a replica.
P-DUR addresses both challenges raised above: certification is deterministic and in some cases avoids synchronization among threads;
when synchronization cannot be avoided, P-DUR implements it efficiently.
P-DUR's key insight is to divide the local database at replicas into \emph{logical partitions} and assign the management of each partition to a different thread/core.
For single-partition transactions, that is, transactions that access data in a single logical partition, execution and termination are handled by the thread responsible for the partition in a manner that resembles classical DUR. 
Single-partition transactions that access different partitions are not mutually synchronized during their execution and termination.
Cross-partition transactions execute concurrently with no synchronization among threads; during termination, all threads involved in the transaction must coordinate in order to certify the transaction.
Threads within a replica coordinate and synchronize by exchanging local messages only; there are no locks set on data items.

We implemented P-DUR and compared its performance and scalability to classical DUR and Berkeley DB. 
Our experiments show that a single P-DUR replica with 16 partitions outperforms a DUR's deployment with 16 replicas by 2.4x and a  multithreaded Berkeley DB server by 10x.
For workloads with single-partition update transactions, P-DUR scales update transactions linearly with the number of partitions. 
Experiments with a social network application confirm P-DUR's scalability and advantage with respect to DUR.

\subsection{Contributions and structure}

This paper makes the following contributions:
First, we revisit deferred update replication and show its inherent performance limitations by means of a simple analytical model inspired by Amdahl's law~\cite{Amdahl}.
Second, we introduce Parallel Deferred Update Replication (P-DUR) and argue that it is fundamentally more efficient than DUR. Notably, under certain workloads, update transactions scale linearly with the number of replicas in P-DUR.
Third, we describe a detailed implementation of P-DUR and compare its performance to DUR and to Berkeley DB, a highly efficient standalone database, using microbenchmarks and a social network application.



The remainder of the paper is structured as follows.
Section~\ref{sec:system_model_and_definitions} introduces our system model and some definitions used in the paper.
Section~\ref{sec:deferred_update_replication} recalls the deferred update replication approach and reasons about its performance limitations.
Section~\ref{sec:parallel_deferred_update_replication} introduces Parallel Deferred Update Replication and models its performance.
Section~\ref{sec:implementation} presents our prototype and some optimizations.
Section~\ref{sec:performance} evaluates P-DUR's performance.
Section~\ref{sec:related-work} reviews related work and Section~\ref{sec:conclusion} concludes the paper.
In the Appendix we argue about P-DUR's correctness.


\section{System model and definitions}
\label{sec:system_model_and_definitions}

We consider a system composed of an unbounded set $C = \{ c_1, c_2, ... \}$ of client processes and a bounded set $S = \{ s_1, ..., s_n\}$ of server processes. 
We assume a benign failure model in which processes can fail by crashing but never perform incorrect actions (i.e., no Byzantine failures). 
Processes have access to stable storage whose state survives failures.
A process, either client or server, that never crashes is \emph{correct}, otherwise the process is \emph{faulty}. 

Processes communicate using either one-to-one or one-to-many communication. 
One-to-one communication uses primitives \emph{send}$(m)$ and \emph{receive}$(m)$, where $m$ is a message.
If both sender and receiver are correct, then every message sent is eventually received. 
One-to-many communication relies on atomic multicast, with primitives \emph{amcast}$(g,m)$ and \emph{deliver}$(g,m)$, where $g$ is a group of processes.
Atomic multicast ensures that 
(a)~a message multicast by a correct process to group $g$ will be delivered by all correct processes in $g$; 
(b)~if a process in $g$ delivers $m$, then all correct processes in $g$ deliver $m$; and
(c)~every two processes in $g$ deliver messages in the same order.
%
The system is asynchronous: there is no bound on messages delays and on relative process speeds.\footnote{Solving atomic multicast requires additional assumptions~\cite{CT96,FLP85}. In the following, we simply assume the existence of an atomic multicast oracle.}

We assume a multiversion database with data items implemented as tuples $\langle k, v, ts \rangle$, where $k$ is a key, $v$ its value, and $ts$ its version.
A transaction is a sequence of read and write operations on data items followed by a commit or an abort operation. 
A transaction $t$ is a tuple $\langle id, st, rs, ws \rangle$ where $id$ is a unique identifier, $st$ is the database snapshot version seen by $t$, $rs$ is the set of data items read by $t$,  $\mathit{readset}(t)$, and $ws$ is the set of data items written by $t$, $\mathit{writeset}(t)$. 
The readset of $t$ contains the keys of the items read by $t$; the writeset of $t$ contains both the keys and the values of the items updated by $t$.
The isolation property is \emph{serializability}: every concurrent execution of committed transactions is equivalent to a serial execution involving the same transactions~\cite{BHG87}.


\section{Deferred Update Replication}
\label{sec:deferred_update_replication}

In the following we briefly review the deferred update replication approach and comment on its performance.

\subsection{Deferred update replication}

In the deferred update replication approach, transactions undergo two phases: the \emph{execution phase}, in which a transaction's read and write operations are processed; and the \emph{termination phase}, in which servers terminate transactions and decide on their outcome, either commit or abort.
Algorithms \ref{alg:dur_client} and \ref{alg:dur_server} illustrate the client and the server sides, respectively, of deferred update replication.

In the execution phase, a client $c$ selects a replica $s$ that will execute $t$'s read operations; $t$'s write operations are buffered by the client until termination. 
Clients keep track of $t$'s readset and writeset (i.e., the keys read and written by $t$). 
During the execution phase, other replicas are not involved in the execution of $t$.

Operation $\mathit{read}(t, k)$ includes key $k$ in $t$'s readset (line~6, Algorithm~1). 
If $t$ has previously updated $k$, the updated value is returned (lines 7 and 8); otherwise the client selects a server $s$ and sends the read request to $s$---this mechanism ensures that transactions read their own writes.
The first read determines $t$'s snapshot, which will be used in all future read operations of $t$. 
Finally, the client returns the value read to the application (lines 10--13). 
Operation $\mathit{write}(t, k, v)$ adds item $(k, v)$ to $t$'s writeset. (lines 14 and 15).

The execution phase of transaction $t$ ends when the client requests to commit $t$. 
Since read operations are guaranteed to see a consistent view of the database, no further steps are required for a read-only transaction to terminate; read-only transactions can commit without certification (lines 17 and 18). 
Clients atomically multicast update transactions to all servers and wait for a response (lines 20--22).

Each server keeps track of variable $\SC$, the latest snapshot that was locally created (line 2, Algorithm 2).
Upon receiving the first read operation for key $k$ from transaction $t$, server $s$ returns the value of $k$ from the latest snapshot $\SC$, together with the value of $\SC$ (lines~3--6), used by the client to set $st$. 
Every subsequent read operation from $t$ will include snapshot $st$ and servers will return a version ``consistent" with $st$. 
A \emph{read on key $k$ is consistent with snapshot} $st$ if it returns the most recent version of $k$ equal to or smaller than $st$ (line 5). This rule guarantees that between the version returned for $k$ and $\SC$ no committed transaction $u$ has modified the value of $k$ (otherwise, $u$'s update would be the most recent one).

Upon delivery of transaction $t$, $s$ first determines the outcome of $t$; 
if $t$ can be committed, $s$ writes $t$'s updates to the local database and replies back to the client (lines 7--11).
The outcome of $t$ is determined by function \emph{certify}, which ensures that $t$ only commits if no transaction $u$, committed after $t$ received its snapshot, updated an item read by $t$ (lines~14--18).

\begin{algorithm}[t]
\small
\caption{Deferred update replication, client $c$'s code}
\label{alg:dur_client}
\begin{distribalgo}[1]
\vspace{1mm}
\INDENT{\textbf{begin$(t)$:}}
	\STATE $t.rs \gets \emptyset$
	\COMMENT{initialize readset}
	\STATE $t.ws \gets \emptyset$
	\COMMENT{initialize writeset}
	\STATE $t.st \gets \bot$
	\COMMENT{initially transactions have no snapshot}
\ENDINDENT
\vspace{1mm}
\INDENT{\textbf{read$(t, k)$:}}
	\STATE $t.rs \gets t.rs \cup \{ k \}$
	\COMMENT{add key to readset}
	\IF[if key previously written...]{$(k, \star) \in t.ws$}
		\RETURN $v$ s.t. $(k,v) \in t.ws$
		\COMMENT{return written value}
	\ELSE[else, if key never written...]
		\STATE send$(\mathrm{read}, k, t.st)$ to some $s \in S$
		\COMMENT{send read request}
		\STATE \textbf{wait until} receive$(k,v,st)$ from $s$
		\COMMENT{wait for response}
		\STATE \textbf{if} $t.st = \bot$ \textbf{then} $t.st \leftarrow st$
		\COMMENT{if first read, init snapshot}
		\RETURN $v$
		\COMMENT{return value from server}
	\ENDIF
\ENDINDENT
\vspace{1mm}
\INDENT{\textbf{write$(t, k, v)$:}}
	\STATE $t.ws \gets t.ws \cup \{(k, v)\}$
	\COMMENT{add key to writeset}
\ENDINDENT
\vspace{1mm}
\INDENT{\textbf{commit$(t)$:}}
	\IF[if transaction $t$ is read-only...]{$t.ws = \emptyset$}
		\RETURN \emph{commit}
		\COMMENT{commit it right away}
	\ELSE[else, if it is an update...]
		\STATE amcast$(S, (c,t))$
		\COMMENT{multicast $t$ to all servers and wait outcome}
		\STATE \textbf{wait until} receive$(\mathit{outcome})$ from $s \in S$
		\COMMENT{ditto}
		\RETURN $\mathit{outcome}$
		\COMMENT{return outcome}
	\ENDIF
\ENDINDENT
\vspace{1mm}
\end{distribalgo}
\end{algorithm}


\begin{algorithm}[t]
\small
\caption{Deferred update replication, server $s$'s code}
\label{alg:dur_server}
\begin{distribalgo}[1]
\vspace{1mm}
\INDENT{\textbf{Initialization:}}
	\STATE $\SC \gets 0$
	\COMMENT{initialize snapshot counter}
\ENDINDENT
\vspace{1mm}
\WHEN{receive$(read,k,st)$ from $c$}
	\STATE \textbf{if} $st = \bot$ \textbf{then} $st \leftarrow \SC$
	\COMMENT{if first read, initialize snapshot}
	\STATE retrieve$(k, v, st)$ from database
	\COMMENT{most recent version $\le st$}
	\STATE send$(k, v, st)$ to $c$
	\COMMENT{return result to client}
\ENDWHEN
\vspace{1mm}
\WHEN{deliver$(c,t)$}
	\STATE $outcome \gets \mbox{certify}(t)$
	\COMMENT{outcome is either commit or abort}
	\IF[if it passes certification...]{$outcome =$ \emph{commit}}
		\STATE apply $t.ws$ to database
		\COMMENT{apply committed updates to db}
	\ENDIF
	\STATE send$(outcome)$ to $c$
	\COMMENT{return outcome to client}
\ENDWHEN
\vspace{1mm}
\FUNCTION[used in line 8]{certify$(t)$}
	\FOR[for all key in readset]{$k \in t.rs$}
		\STATE retrieve$(k, v, st)$ from database
		\IF[if newer version of k exists]{$st > t.st$}
			\RETURN \emph{abort}
			\COMMENT{transaction must abort}
		\ENDIF
	\ENDFOR
	\STATE $\SC \gets \SC + 1$
	\COMMENT{create one more snapshot}
	\RETURN \emph{commit}
	\COMMENT{transaction must commit}
\ENDFUNC
\vspace{1mm}
\end{distribalgo}
\end{algorithm}

\subsection{Performance analysis}
\label{sec:performance_issues}

Since the execution and termination of read-only transactions are entirely local to a server, the performance (i.e., throughput) of read-only transactions grows proportionally with the number of deployed servers.
The same observation does not hold for update transactions, however, since every (correct) replica must be involved in the termination of each update transaction.
Intuitively, the throughput of update transactions is not proportional to the number of replicas.
%

A simple analysis, inspired by Amdahl's law~\cite{Amdahl}, helps illustrate the inherent performance limitations of deferred update replication under update transactions.
Let $\tau_{(1)}$ be the peak throughput (in transactions per second) of DUR when clients submit update transactions to a single replica; the other replicas are used for redundancy only.
If $\gamma_e$ and $\gamma_t$ are the cost (e.g., in operations) needed by the replica to execute and terminate a transaction, respectively, then the corresponding load at the replica (in operations per second) is $\tau_{(1)}(\gamma_e+\gamma_t)$.

Consider now a configuration where $n$ replicas execute transactions and each replica executes approximately the same number of transactions per time unit.
If $\tau_{(n)}$ is the peak throughput of the ensemble, the load at each replica is $\tau_{(n)}(\gamma_e/n+\gamma_t)$, since the execution of transactions is shared among replicas, but not their termination.

Since the number of operations per time unit that a replica can execute is a property of the replica and does not depend on the number of replicas:
\begin{equation}
\tau_{(1)}(\gamma_e+\gamma_t) = \tau_{(n)}(\gamma_e/n+\gamma_t)
\end{equation}
which can be rearranged as
\begin{equation}
\label{eq2}
\tau_{(n)} = \tau_{(1)}n(\gamma_e+\gamma_t)/(\gamma_e+n\gamma_t).
\end{equation}

Equation (\ref{eq2}) relates the throughput of DUR when a single replica executes transactions and when $n$ replicas execute transactions.
We can derive from this relation DUR's \emph{scaling}, which captures the inherent performance limitations of the approach:
\begin{equation}
\mathcal{S}_{\textit{DUR}}(n) = \frac{\tau_{(n)}}{\tau_{(1)}} = \frac{n(\gamma_e+\gamma_t)}{\gamma_e+n\gamma_t}
\end{equation}

Ideally, throughput would be directly proportional to the number of replicas (i.e., $\mathcal{S}_{\textit{DUR}}(n)$ would be $n$).
%
This is the case of read-only transactions, where $\gamma_t=0$.
For update transactions, however, augmenting the number of replicas does not increase throughput in the same proportion.
%
We can determine the maximum performance one can expect from update transactions in DUR as follows.  
\begin{equation}
\label{eq4}
\mathcal{S}_{\textit{DUR}}(\infty) = \lim_{n\to\infty}\frac{n(\gamma_e+\gamma_t)}{\gamma_e+n\gamma_t}=\frac{\gamma_e+\gamma_t}{\gamma_t}.
\end{equation}

%

Replication (scaling out) provides limited performance improvements to update transactions in deferred update replication, although it boosts system availability.
In the next section, we propose a technique to parallelize the execution and termination of update transactions in a server (scaling up).


\section{Parallel Deferred Update Replication}
\label{sec:parallel_deferred_update_replication}

In this section, we extend the deferred update replication approach to account for multi-core architectures.
The goal is to allow both the execution and the termination of transactions to proceed in parallel. 
Our key insight is to divide the database into logical partitions and let threads execute sequentially within a partition and concurrently across partitions.
In the following, we revisit our system model, introduce the general idea of parallel deferred update replication, discuss the details of the approach, and analyze its performance using an analytical model.
In the Appendix we reason about P-DUR's correctness.

\subsection{Additional definitions and assumptions}
\label{sub:additional_definitions_and_assumptions}

We divide the database into $P$ \emph{logical partitions} such that each key $k$ belongs to a single partition, which we denote $\mathit{partition}(k)$. 
We distinguish between two different types of transactions: \emph{single-partition} and \emph{cross-partition}. 
The execution and termination of single-partition transactions involve one partition only; cross-partition transactions involve two or more partitions. 
We denote $\mathit{partitions}(t)$ the set of partitions that contain items read or written by transaction $t$.

We refine the notion of a server $s_i$ to be composed of $P$ processes, $s_i = \{ p_{i,1}, ..., p_{i,P} \}$, as opposed to a single process as defined in Section~\ref{sec:system_model_and_definitions}.
For simplicity, we assume that all processes in a faulty server fail simultaneously; processes in a correct server never fail.
Processes within a server communicate using the same primitives defined in Section~\ref{sec:system_model_and_definitions}; in practice, communication among processes in the same server is more efficient than communication between processes across servers.

%

\subsection{General idea}
\label{sub:general_idea}

\begin{figure*}[ht]
	\begin{center}
		\includegraphics[width=0.7\textwidth]{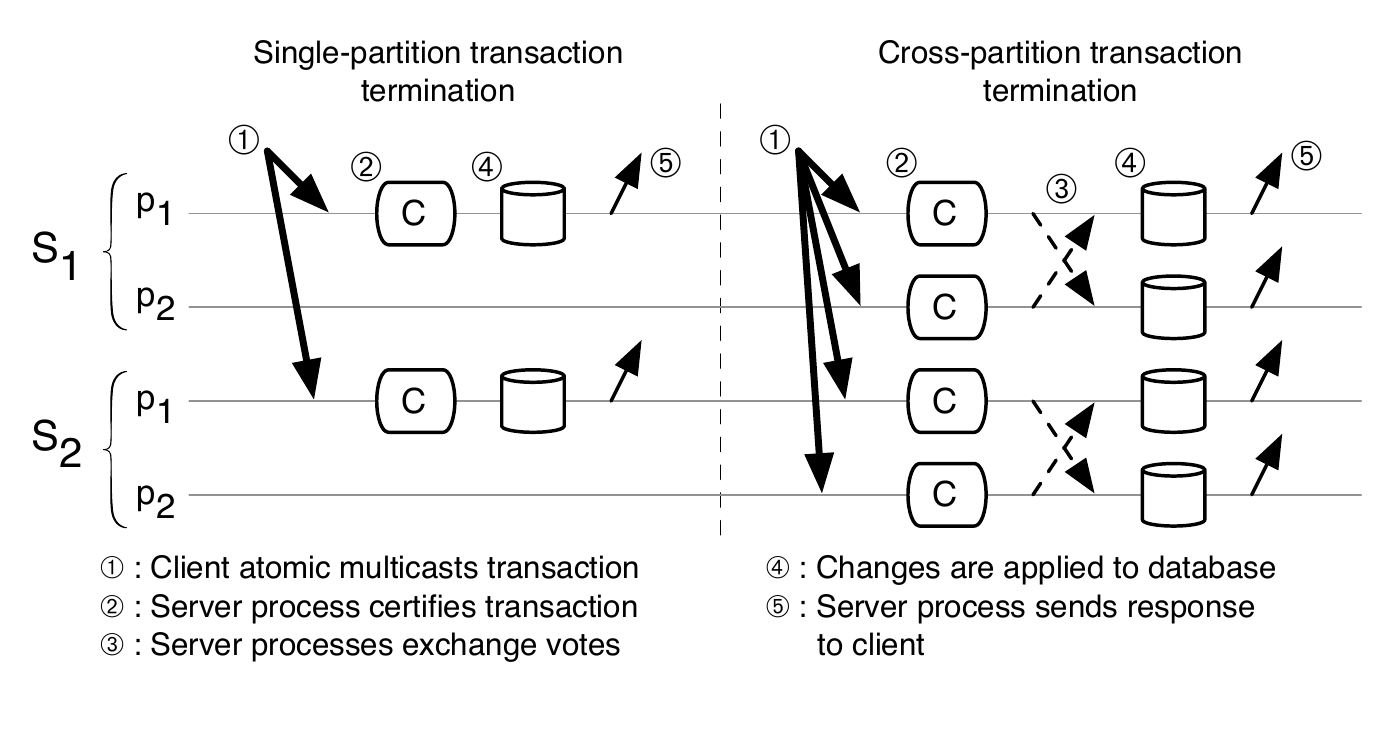} 
		\caption{\small Termination of single- and cross-partition transactions in Parallel-DUR. (Legend: thick arrows are multicast communication; regular arrows are process communication across nodes; dashed arrows are process communication within a server.)}
		\label{fig:parallel-dur}
	\end{center}
\end{figure*}

In P-DUR, each server hosts logical partitions, each one under the control of an independent process.
Single-partition transactions are handled in the same way as in the classical deferred update replication approach. 
Different than DUR, however, single-partition transactions that access different logical partitions can execute and terminate concurrently.
Cross-partition transactions execute independently in two or more logical partitions; at termination, processes at the concerned partitions certify the transaction with respect to other transactions that executed in the same partition and exchange their votes.
The transaction is committed if it passes certification at all concerned partitions.
Figure \ref{fig:parallel-dur} illustrates the termination of single-partition and cross-partition transactions.

Two characteristics account for the good performance of cross-partition transactions in P-DUR.
First, the termination of cross-partition transactions undergoes a two-phase commit-like protocol where votes are exchanged among processes within the same server.
Communication within a server is more efficient than communication across servers.
Second, atomic multicast's order property ensures that transactions that access common partitions are delivered  in the same order by the common partitions (within a server and across servers).
Therefore, once a partition casts a vote for a transaction, it does not need to keep locks on the values read by the transaction (to ensure the validity of the certification test until the transaction commit), resulting in a deadlock-free procedure.

\subsection{Algorithm in detail}
\label{sub:algorithm_in_detail}

Algorithms \ref{alg3} and \ref{alg4} illustrate the client and the server sides of Parallel Deferred Update Replication.
Each client keeps a vector of snapshots, one per logical partition (line~4 in Algorithm~\ref{alg3}).
To execute a read operation on a key, the client first checks whether the key has been previously written (line~7).
If so, then the value written is returned (line~8); otherwise, the client determines the logical partition the key belongs to (line~10) and sends the read request to the specified partition in some server $s$ (line~11).
The client eventually receives the value read within a snapshot (line~12), which the client stores for future reads within the same partition (line~13).
For brevity, Algorithm~\ref{alg3} does not handle the failure of $s$.
Upon suspecting the crash of $s$, the client can contact another server.
As in DUR, writes are buffered by the client (lines~15--16).
To commit transaction $t$, the client atomically multicasts $t$ to partitions read and written by $t$ in all servers (line~18).

Servers handle read requests in P-DUR and in DUR in the same way.
When a process $p$ in server $s$ delivers a transaction for termination (line~7 in Algorithm~\ref{alg4}), it certifies $t$ based on $p$'s logical partition.
If $t$ is a cross-partition transaction, $p$'s vote for $t$ is sent to the other processes involved in $t$ (line~11), and $p$ waits for all votes for $t$ (line~12). More precisely, $p$ only needs to send a vote to all local processes serving the partitions involved (in the same server), and similarly $p$ waits only for one vote from the each local partition involved by $t$. One vote from each partition is sufficient, since different replicas of the same partition compute the votes deterministically. Vote exchanges can be efficiently implemented as it involves only local communication (i.e., interprocess communication).

If at least one vote is to abort $t$, $t$ will be aborted (lines~13--14).
If $t$ is single-partition, its fate depends on the result of $p$'s certification alone.
If $t$ passes certification (line~15), $t$'s writes are applied to the local database (line~16).
In any case, the client is notified of $t$'s outcome (line~17).
Transaction certification in P-DUR and DUR are similar.


\begin{algorithm}[t]
\small
\caption{Parallel DUR, client $c$'s code}\label{alg3}
\begin{distribalgo}[1]
\vspace{1mm}
\INDENT{\textbf{begin$(t)$:}}
	\STATE $t.rs \gets \emptyset$
	\COMMENT{initialize readset}
	\STATE $t.ws \gets \emptyset$
	\COMMENT{initialize writeset}
	\STATE $t.st[1 ... P] \gets [\bot ... \bot]$
	\COMMENT{initialize vector of snapshot times}
\ENDINDENT
\vspace{1mm}
\INDENT{\textbf{read$(t, k)$:}}
	\STATE $t.rs \gets t.rs \cup \{ k \}$
	\COMMENT{add key to readset}
	\IF[if key previously written...]{$(k, \star) \in t.ws$}
		\RETURN $v$ s.t. $(k,v) \in t.ws$
		\COMMENT{return written value}
	\ELSE[else, if key never written...]
		\STATE $q \gets \mathit{partition}(k)$
		\COMMENT{get the key's partition}
		\STATE send$(\mathrm{read}, k, t.st[p])$ to process $p_{s,q}$ for some server $s$
		\STATE \textbf{wait until} receive $(k,v,st)$ from $p_{s,q}$
		\COMMENT{wait response}
		\STATE \textbf{if} $t.st[q]\!=\!\bot$ \textbf{then} $t.st[q]\!\leftarrow\!st$
		\COMMENT{if first read, init snapshot}
		\RETURN $v$
		\COMMENT{return value from server}
	\ENDIF
\ENDINDENT
\vspace{1mm}
\INDENT{\textbf{write$(t, k, v)$:}}
	\STATE $t.ws \gets t.ws \cup \{(k, v)\}$
	\COMMENT{add key to writeset}
\ENDINDENT
\vspace{1mm}
\INDENT{\textbf{commit$(t)$:}}
	\STATE $\mathit{amcast}(\mathit{partitions}(t), (c,t))$
	\STATE \textbf{wait until} $receive(\mathit{outcome})$ from $p$
	\RETURN \emph{outcome}
	\COMMENT{outcome is either commit or abort}
\ENDINDENT
\vspace{1mm}
\end{distribalgo}
\end{algorithm}


\begin{algorithm}[]
\small
\caption {Parallel DUR, server $s$'s code in partition $p$}\label{alg4}
\begin{distribalgo}[1]
\vspace{1mm}
\INDENT{\textbf{Initialization:}}
	\STATE $\SC \gets 0$
	\COMMENT{snapshot counter}
\ENDINDENT
\vspace{1mm}
\WHEN{receive$(read,k,st)$ from $c$}
	\STATE \textbf{if} $st\!=\!\bot$ \textbf{then} $st\!\leftarrow\!\SC$
	\COMMENT{if first read, init snapshot}
	\STATE retrieve$(k,v,st)$ from database
	\COMMENT {most recent version $\le st$}
	\STATE send$(k, v, st)$ to $c$
	\COMMENT {return result to client}
\ENDWHEN
\vspace{1mm}
\WHEN{deliver$(c, t)$}
	\STATE $\mathit{v} \gets \mbox{certify}(t)$
	\COMMENT{compute the local outcome}
	\IF{$t$ is cross-partition}
		\STATE $Q \gets \{ p_{s,r} : r \in \mathit{partitions}(t) \}$
		\STATE $\mbox{send}(t.id, p, v)$ to all processes in $Q$
		\STATE \textbf{wait until} receive$(t.id, q, v)$ from processes in $Q$
		\IF{one or more votes contain \emph{abort}}
			\STATE $v \gets \mathit{abort}$
		\ENDIF
	\ENDIF
	\IF[if transactions commits]{$\mathit{v} = \mathit{commit}$}
		\STATE apply $t.ws$ with version $\SC$ to database
	\ENDIF
	\STATE $\mbox{send}(\mathit{outcome})$ to c
	\COMMENT{return outcome to client}
\ENDWHEN
\vspace{1mm}
\FUNCTION[used in line 9]{certify$(t)$}
	\FOR[for all key in readset]{$k \in t.rs$}
		\STATE retrieve$(k, v, st)$ from database
		\IF[if newer version of k exists]{$st > t.st[p]$}
			\RETURN \emph{abort}
			\COMMENT{transaction must abort}
		\ENDIF
	\ENDFOR
	\STATE $\SC \gets \SC + 1$
	\COMMENT{create one more snapshot}
	\RETURN \emph{commit}
	\COMMENT{transaction must commit}
\ENDFUNC
\vspace{1mm}
\end{distribalgo}
\end{algorithm}

\subsection{Performance analysis}
\label{sec:p-dur_perf_issues}

We now derive a performance model for P-DUR.
Recall from Section~\ref{sec:performance_issues} that $\gamma_e$ and $\gamma_t$ are the cost (in operations) for a replica to execute and terminate a transaction, respectively.

Let $\tau_{(n,p,g)}$ be the peak throughput of a configuration of P-DUR where $n$ is the number of replicas that execute and terminate transactions, $p$ is the number of partitions, and $g$ the percentage of cross-partition transactions in the workload (i.e., a number between 0 and 1).
Similarly to DUR's analysis, we assume that each replica executes approximately the same number of transactions per time unit. 
In the case of P-DUR, this assumption implies that transactions are equally distributed across partitions.
For simplicity, assume that cross-partition transactions involve all $p$ partitions in the configuration.

The load in operations per time unit executed by one replica with a single partition is $\tau_{(1,1,1)}(\gamma_e+\gamma_t)$.
With $n$ replicas and $p$ partitions, all partitions in one replica execute cross-partition transactions but only one partition in a replica executes single-partition transactions; all partitions in all replicas must terminate cross-partition transactions, but only one partition in each replica terminates single-partition transactions; therefore, the load at a replica is
$\tau_{(n,p,g)}(g\gamma_e/n + (1-g)\gamma_e/(np) + g\gamma_t +(1-g)\gamma_t/p)$.

Since the number of operations per time unit that a replica can execute is a property of the replica, we can determine P-DUR's scaling:

\begin{equation}
\mathcal{S}_{\textit{P-DUR}}(n,p,g) = \frac{\tau_{(n,p,g)}}{\tau_{(1,1,1)}} = \frac{np(\gamma_e+\gamma_t)}{(\gamma_e+n\gamma_t)(1-g+pg)}
\end{equation}

By considering P-DUR's and DUR's scaling we can get a few insights about how the two approaches compare.


\emph{How does P-DUR compare to DUR when we increase the number of replicas arbitrarily in both approaches?}
We determine P-DUR's scaling with an arbitrarily large number of replicas in the best case, when transactions are local, and in the worst case, when transactions are cross-partition:
\begin{equation}
\mathcal{S}_{\textit{P-DUR}}(\infty,p,0) = \frac{p(\gamma_e+\gamma_t)}{\gamma_t} = p\ \mathcal{S}_{\textit{DUR}}(\infty)
\end{equation}
\begin{equation}
\mathcal{S}_{\textit{P-DUR}}(\infty,p,1) = \frac{\gamma_e+\gamma_t}{\gamma_t} = \mathcal{S}_{\textit{DUR}}(\infty)
\end{equation}
Equation (6) shows that when transactions are single-partition, each P-DUR partition behaves as an independent instance of DUR: $p$ partitions improve P-DUR's performance over DUR by $p$ times.
When all transactions are cross-partition, however, P-DUR behaves just like DUR (equation (7)).

\emph{How do scaling up and scaling out compare?}
We address this issue by comparing the scaling of a single replica in P-DUR (scaling up) and the scaling of DUR (scaling out):
\begin{equation}
\mathcal{S}_{\textit{P-DUR}}(1,\infty,g) = \lim_{p\to\infty} \frac{p}{pg-g+1} = \frac{1}{g}\\
\end{equation}
\begin{equation}
\frac{\mathcal{S}_{\textit{P-DUR}}(1,\infty,g)}{\mathcal{S}_{\textit{DUR}}(\infty)} = \frac{\gamma_t}{g(\gamma_e+\gamma_t)}
\end{equation}

From equation (9), P-DUR's scaling (scaling up) is greater than DUR's
scaling (scaling out) when $\gamma_t/g(\gamma_e+\gamma_t)>1$, which
can be rearranged as $g<\gamma_t/(\gamma_e+\gamma_t)$.  For example,
if $\gamma_e=\gamma_t$, we can expect that increasing the number of
partitions in P-DUR will result in greater throughput than increasing
the number of replicas in DUR when fewer than 50\% of P-DUR's
transactions are cross-partition. 


Notice that while our models allow us to compare the inherent limitations of P-DUR and DUR, they do not capture several important aspects of the two techniques.
For example, we assume that the cost to execute and terminate transactions is fixed (i.e., $\gamma_e$ and $\gamma_t$), while in reality these costs may vary, depending on various optimizations (e.g., caching).
The termination of cross-partition transactions in P-DUR involves steps we do not model, such as the exchange of votes across partitions.
Finally, cross-partition transactions do not necessarily access all partitions, as we have assumed.
In Section~\ref{sec:performance} we compare P-DUR and DUR experimentally using benchmarks that account for these aspects.


%
%
%

\subsection{Parallel vs.\ Distributed DUR}
\label{sub:pdur_sdur}

P-DUR starts with a replicated database (DUR) and partitions each replica (scaling up).
An alternative approach is to scale out the system by partitioning the database first and then distributing replicas of each partition across many nodes.
One example of this approach in the context of DUR was proposed by Sciascia et al. in~\cite{SPJ12}, hereafter referred to as Distributed DUR.
While one can achieve scalability of update transactions with both approaches, they differ in important aspects.
On the one hand, P-DUR can optimize cross-partition transactions since communication across partitions happens within a single node and partitions can be assumed to fail as a single unit (e.g., if each partition is implemented as a thread within the same operating system process).
On the other hand, Distributed DUR can accommodate large databases that would not fit in the main memory of a single node.
Moreover, it allows to configure replication on a partition basis.
This is important for performance since read-only transactions in a partition scale with the number of partition replicas and therefore one could allocate replicas based on the rate of read-only transactions expected within the partition (i.e., ``hot spot partitions" could have more replicas than partitions rarely read).
One interesting direction for future work is to consider a hybrid technique combining database parallelism and distribution in DUR.


\section{Implementation}
\label{sec:implementation}

Our P-DUR prototype is fully implemented in C.
Our implementation matches most of Algorithm 4 and slightly differs from it in a few aspects.
In the following, we report the most significant changes we made in our prototype.

Clients connect to a single process only, and submit all their read requests to that process. 
If a process receives a request for a key $k$ it does not store, the process fetches the missing entry from the process serving partition $k$.
The partitioning of keys is therefore transparent to the clients.
We use TCP for all the communication across servers and Unix domain sockets for interprocess communication (i.e., for exchanging votes and fetching keys from other partitions).

Instead of using an atomic multicast, as shown in Algorithm 4, our
implementation uses independent instances of atomic broadcast, each
instance associated with a logical partition. Single-partition
transactions are atomically broadcast to the processes associated with
the involved partition.  Cross-partition transactions are broadcast to
each partition independently (i.e., one broadcast per partition, all
executed in parallel).  As a consequence, partitions might deliver two
cross-partition transactions, say $t_1$ and $t_2$, in different order
(e.g., process $p$ in one partition delivers $t_1$ and then $t_2$ and
process $q$ in another partition delivers $t_2$ and then $t_1$).  We
employ a stronger certification test to guarantee serializability
despite out-of-order delivery across partitions.  The stronger
certification test ensures that transactions can be serialized in any
order (i.e., $t_1$ before $t_2$, and $t_2$ before $t_1$).  To do so
the certification test checks for intersections between the readset of
$t_1$ against the writeset of $t_2$, and vice-versa. 
Atomic broadcast within processes in the same partition was implemented with Paxos.\footnote{https://github.com/sambenz/URingPaxos}





\section{Performance}
\label{sec:performance}

In this section, we evaluate the performance of P-DUR and compare it to classical DUR and Berkeley DB, a popular standalone database.
We first detail the benchmarks used and the experimental setup, and then discuss our results.

\subsection{Benchmarks}

We used two different benchmarks to assess the performance of deferred update replication: a microbenchmark and a Twitter-like social network application.

Table~\ref{tbl:bench} shows all transaction types used in our microbenchmark experiments.
%
Transaction type I characterizes short transactions that make few changes to the database;
transaction type II was designed to stress the execution and termination of transactions (recall that transaction execution issues essentially read operations and termination must check the validity of each value read);
finally, transaction type III strikes a balance between reads and writes while stressing database updates. 
In the microbenchmarks, before collecting results, the database was populated with 4.2 million entries.

\begin{table}[htdp]
\centering
\begin{tabular}{c|c|c|c|c|c}
Type & Reads & Writes & Key size & Value size & DB size \\
        & (ops) & (ops)  & (bytes)  & (bytes)    & (items) \\
\hline
I   & 2  & 2  & 4 & 4 & 4.2M \\
II  & 32 & 2  & 4 & 4 & 4.2M \\
III & 16 & 16 & 4 & 4 & 4.2M \\
\end{tabular}
\caption{Transaction types in microbenchmark.}
\label{tbl:bench}
\end{table}%

\clearpage
The Twitter-like benchmark implements the typical operations of a social networking application. Usually the users of such application can: follow other users; post messages; and retrieve their timeline containing the messages of users they follow.
The benchmark works as follows. For each user $u$, the database keeps track of: a list of ``consumers'' which contains the user ids that follow $u$; a list of ``producers'' which contains user ids that $u$ follows; and $u$'s list of posts.
In the experiments we partitioned the database by user. Each user account is stored together with its posts, producer list, and consumer list.

A post transaction consists in appending a new message to the list of
posts. Given our partitioning, post transactions are
single-partition. Follow transactions update two lists, a consumer
list and a producer list of two different users. Follow transactions
are single-partition if both user accounts are stored in the same
partition. Or cross-partition otherwise. Timeline transactions build a
timeline for user $u$ by merging the last posts of $u$'s
producers. Timeline transactions are cross-partition and read-only.

\subsection{Evaluation setup}

In some of our experiments, we compared the performance of P-DUR with Berkeley DB\footnote{http://www.oracle.com/technetwork/products/berkeleydb} version 5.2 (hereafter BDB), an open-source library for embedded databases. We setup BDB to use the in-memory B-Tree access method with transactions enabled. To make the comparison fair, we spawn several client threads issuing RPC requests against a server that embeds BDB. We used Apache Thrift\footnote{http://thrift.apache.org/} to implement an efficient multithreaded RPC server.

In the experiments we used a node running all server processes accepting client requests. 
In P-DUR this server hosts a single replica with multiple partitions and in DUR the server hosts multiple independent replicas.
The server is an HP ProLiant DL165
G7 node, equipped with two eight-core AMD Opteron(TM) 6212 processors
running at 2.6GHz and 128GB of memory, connected to a HP ProCurve
2910al-48G gigabit switch. 

Clients ran on Dell PowerEdge 1435 nodes equipped with two dual-core
AMD Opteron(tm) 2212 processors running at 2GHz and 4 GB of main
memory. Client nodes are connected to an HP ProCurve Switch 2900-48G
gigabit network switch. The switches are linked by two 10Gbps
links. The round-trip time between clients and the server is 0.19 ms
for a 1KB packet.

Each Paxos instance uses three acceptors with in-memory storage. 
Paxos acceptors ran on HP ProLiant DL160 G5 nodes
equipped with two quad-core Xeon L5420 @ 2.5GHz processors and 8 GB of
main memory. These nodes are connected to the same switch as the
server machine. All nodes ran CentOS Linux 6.3 64-bit with kernel
2.6.32.

In the following sections, we report the highest throughput achieved in each configuration and the 90-th percentile latency that corresponds to the maximum throughput.

\subsection{Baseline performance}
\label{sec:baseline}


%
In these experiments, we measure throughput and latency as we increase the processing capacity of the various approaches.
In P-DUR we increase the number of logical partitions (cores); in DUR we increase the number of replicas since each replica is single-threaded, as explained previously in the text;
BDB runs as a standalone server where we control the number of concurrent threads receiving client requests.

\begin{figure*}[ht]
  \begin{center}
    \begin{tabular}{cc}
      \includegraphics[width=\sizefactor\columnwidth]{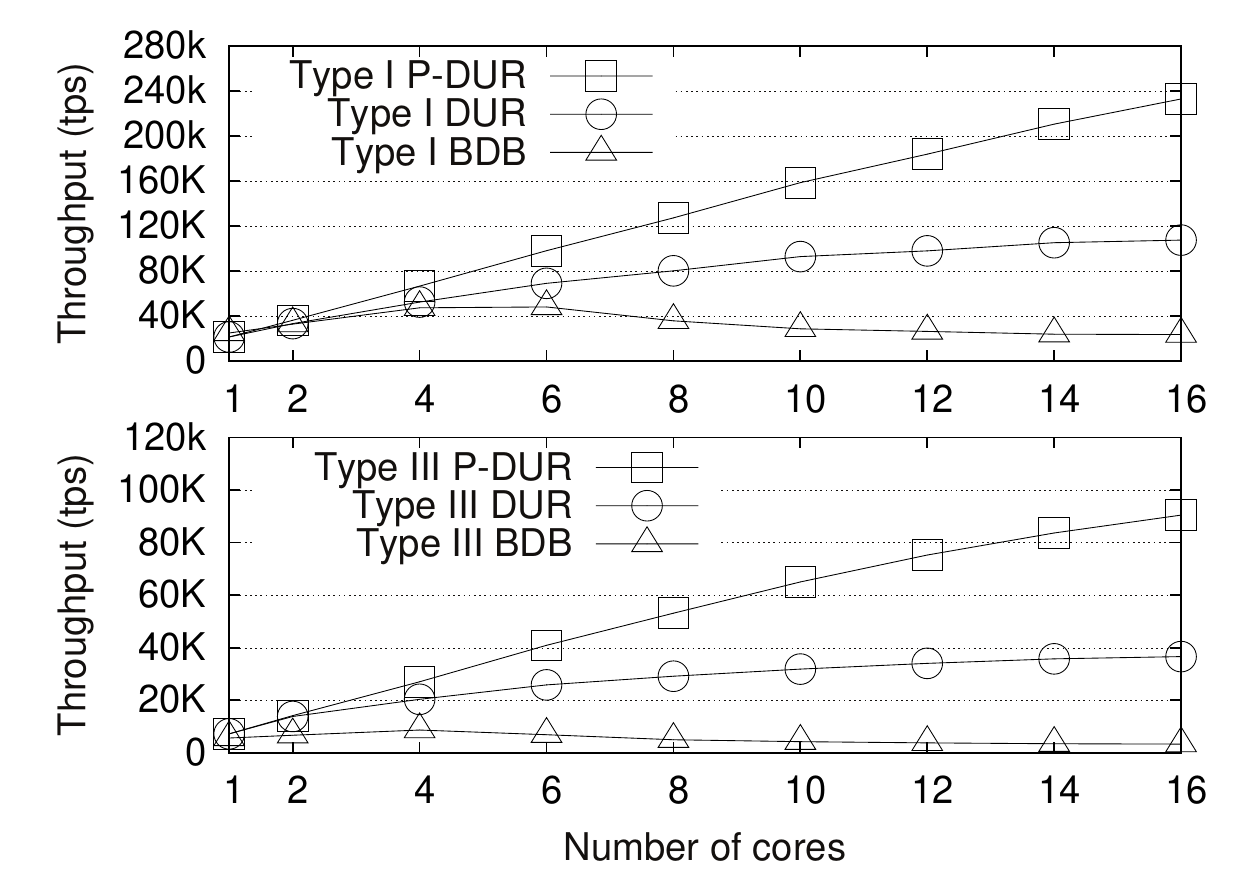} 
      \includegraphics[width=\sizefactor\columnwidth]{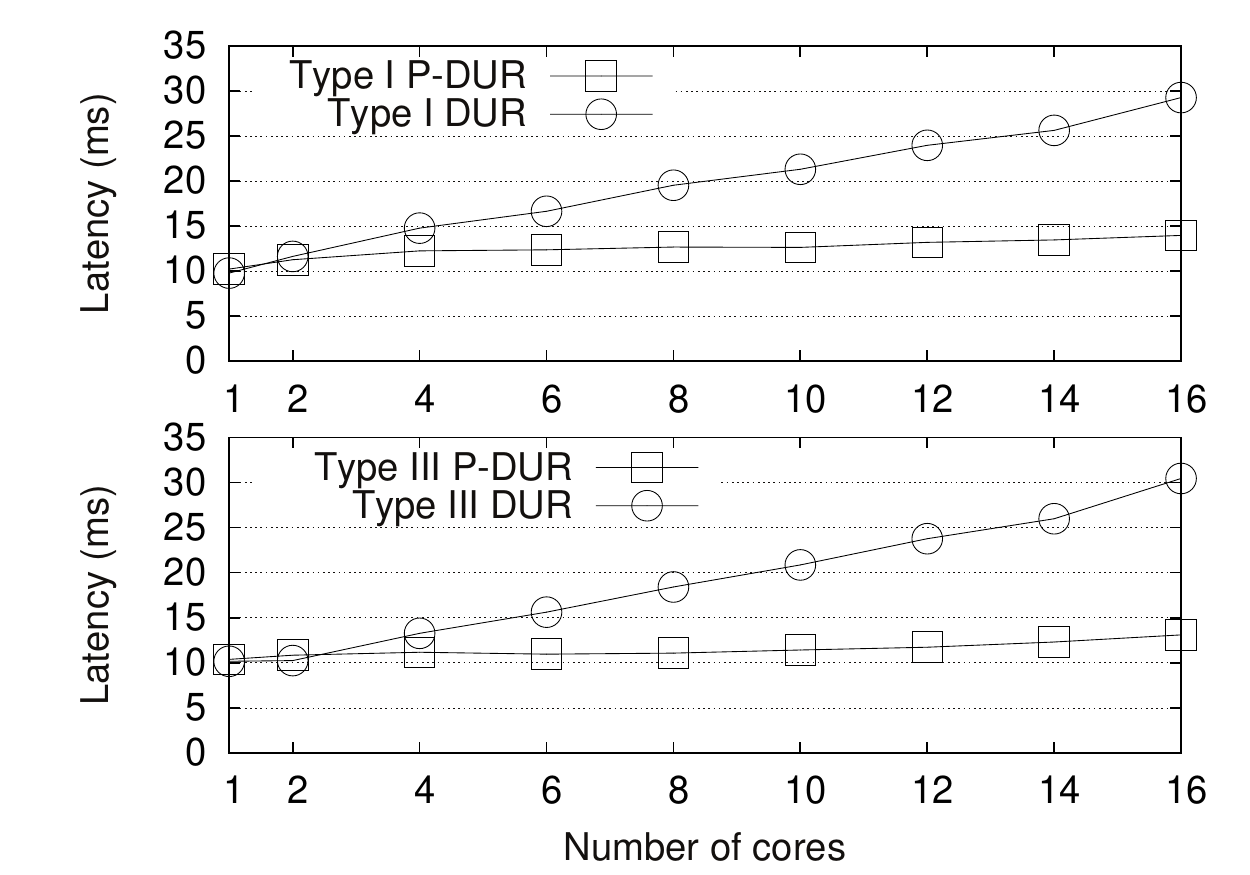} \\
    \end{tabular}
    \caption{\small P-DUR versus DUR and BDB. In P-DUR and BDB a single node executes transactions and the number of cores in the node varies; in DUR we vary the number of replicas.}
    \label{pdurthr}
  \end{center}
	\vspace{-2mm}
\end{figure*}

%
Figure~\ref{pdurthr} focuses on the results for transactions of type I and III. 
We have also run experiments with transactions type II but omit the results since they are very similar to type III's results.
We draw the following conclusions from the experiments. 
First, P-DUR's throughput increases almost
proportionally to the number of cores up to 16 cores, the number of
cores of the machine we used. Second, the performance of DUR is the
same as P-DUR's when both are given 1 core, as expected. DUR's
performance degrades quickly as we increase the number of
replicas. Measurements of CPU utilization showed that DUR's replicas
are CPU-bound, consuming close to 100\% of the CPU at 16 replicas.
Third, BDB benefits from multiple cores up to 4 cores; additional
cores resulted in a degradation of performance, a behavior previously observed in the literature~\cite{Johnson:2009}. Similar behavior was
observed for both transaction type I and III.  
Finally, DUR's latency
grows quickly as we increase the number of replicas, suggesting that transactions
tend to queue up due to CPU congestion. The latency of P-DUR
increases slowly past 4 cores.

\subsection{Scalability of P-DUR versus DUR}
\label{sec:scalability}

\begin{figure*}[ht]
  \begin{center}
    \begin{tabular}{cc}
      \includegraphics[width=\sizefactor\columnwidth]{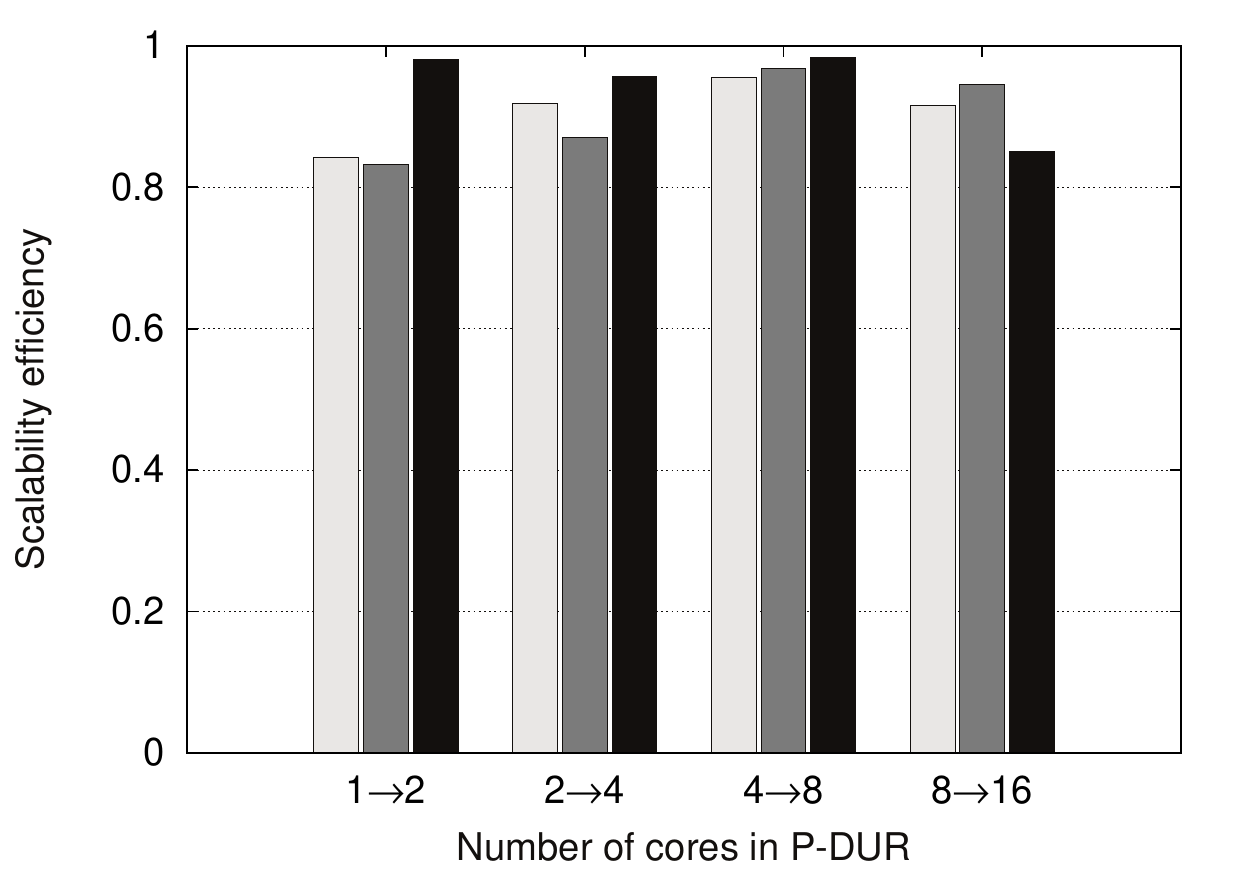} 
      \includegraphics[width=\sizefactor\columnwidth]{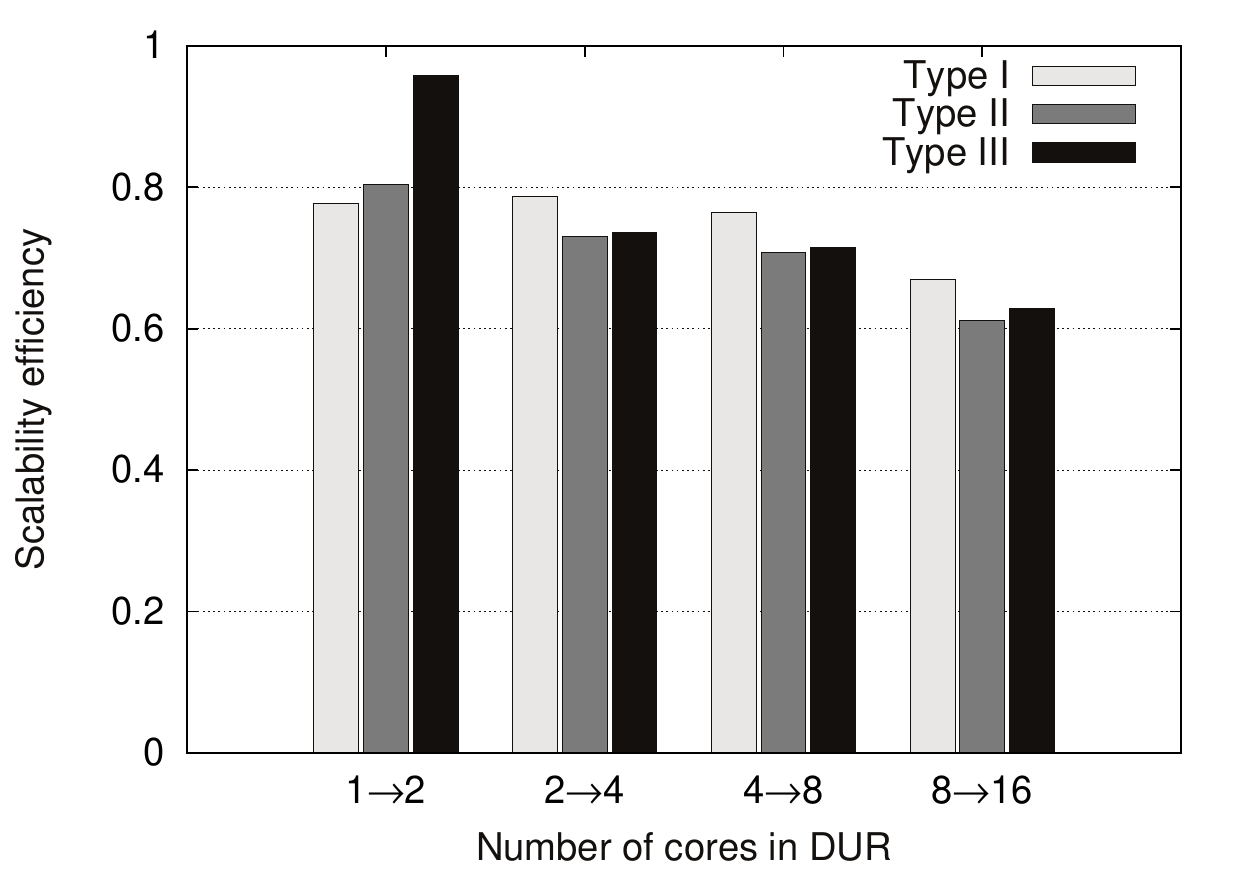} \\
    \end{tabular}
    \caption{\small Throughput and latency of P-DUR versus DUR.}
    \label{dur}
  \end{center}
  \vspace{-2mm}
\end{figure*}

%
Figure~\ref{dur} reports the \emph{scalability efficiency} \cite{sinfonia} when increasing the
number of partitions/cores in P-DUR (left) and the number of replicas in DUR (right). In the scalability
efficiency graph, each bar shows how the throughput changes when
doubling the number of partitions and replicas. A scalability efficiency of 1 indicates
that the system does not experience any degradation when increasing in size from $n$ to $2n$, in other words, it scales perfectly.

%
P-DUR maintains scalability efficiency relatively stable between 83\% (type II, from 1 to 2 partitions) and 98\% (type III, from 4 to 8 partitions) for all transaction types.
This suggests that if additional resources were available at the replica (i.e., cores), throughput would scale further with the added resources.
With two exceptions, DUR's scalability efficiency is below 80\% for all transaction types.
More limiting, however, is the degradation trend experienced by DUR as the number of replicas doubles.


\subsection{The effect of cross-partition transactions}
\label{sec:crosstxns}

\begin{figure*}[ht]
  \begin{center}
    \begin{tabular}{cc}
      \includegraphics[width=\sizefactor\columnwidth]{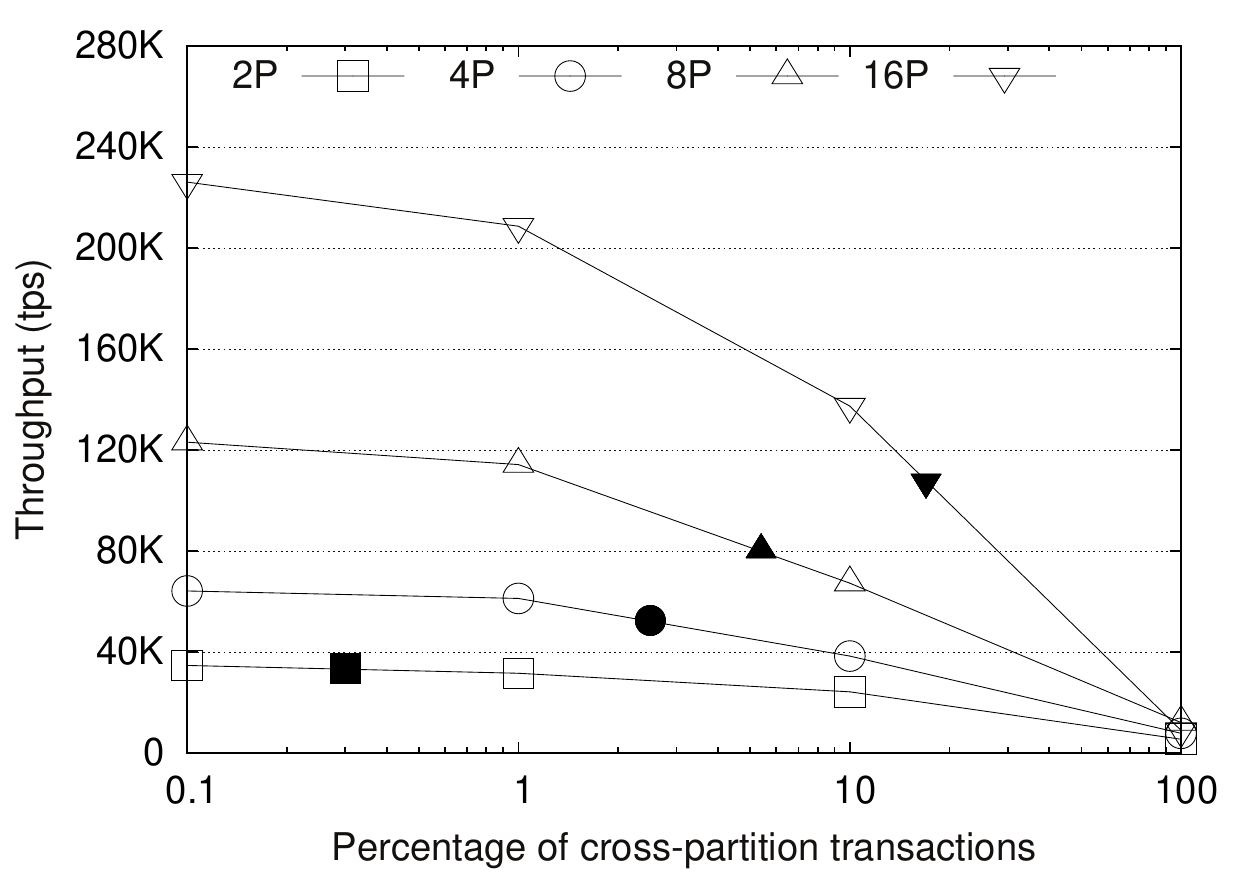} 
      \includegraphics[width=\sizefactor\columnwidth]{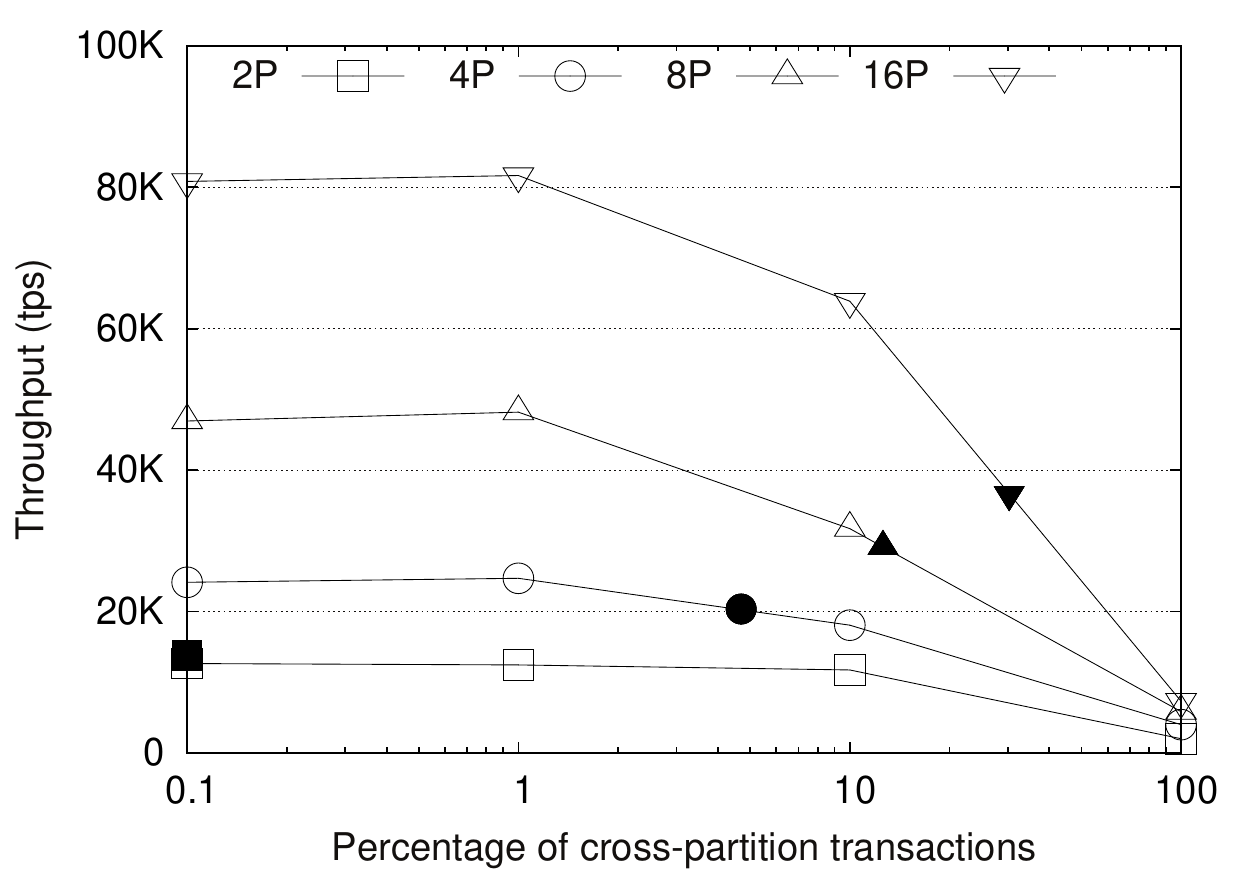} \\
    \end{tabular}
    \caption{\small Throughput of P-DUR with cross-partition transactions of Type I (left) and Type III (right).}
    \label{xpart}
  \end{center}
	\vspace{-2mm}
\end{figure*}

%
The experiments in Figure~\ref{xpart} assess how workloads with different percentages of cross-partition transactions affect P-DUR's throughput. 
We run experiments with transaction types I (left) and III (right). 
Each cross-partition transaction accesses two partitions, generated randomly. 
We vary the percentage of cross-partition transactions in the workload from 0.1\% to 100\%.

%
As expected, cross-partition transactions have a negative impact on throughput since their termination requires additional operations and coordination among threads. 
The black points in the figure represent the throughput of DUR using as many replicas as the number of partitions in P-DUR. 
These points show the percentage of cross-partition transactions after which DUR
has the advantage in a given configuration. The figure shows that for
small configurations, most of the time DUR outperforms P-DUR, given
even a small percentage of cross-partition transactions. 
However, as the system size
grows, P-DUR's scalability pays off and it compensates the overhead for larger
percentages of cross-partition transactions. 
These results show that properly partitioning the dataset to avoid cross-partition transactions is important for performance.
The problem of how to partition the database for performance is an active area of current research (e.g.,~\cite{curino2010schism, pavlo2012skew}).

\subsection{Social network application}
\label{sec:twitterperf}

\begin{figure*}[ht]
  \begin{center}
    \begin{tabular}{cc}
      \includegraphics[width=\sizefactor\columnwidth]{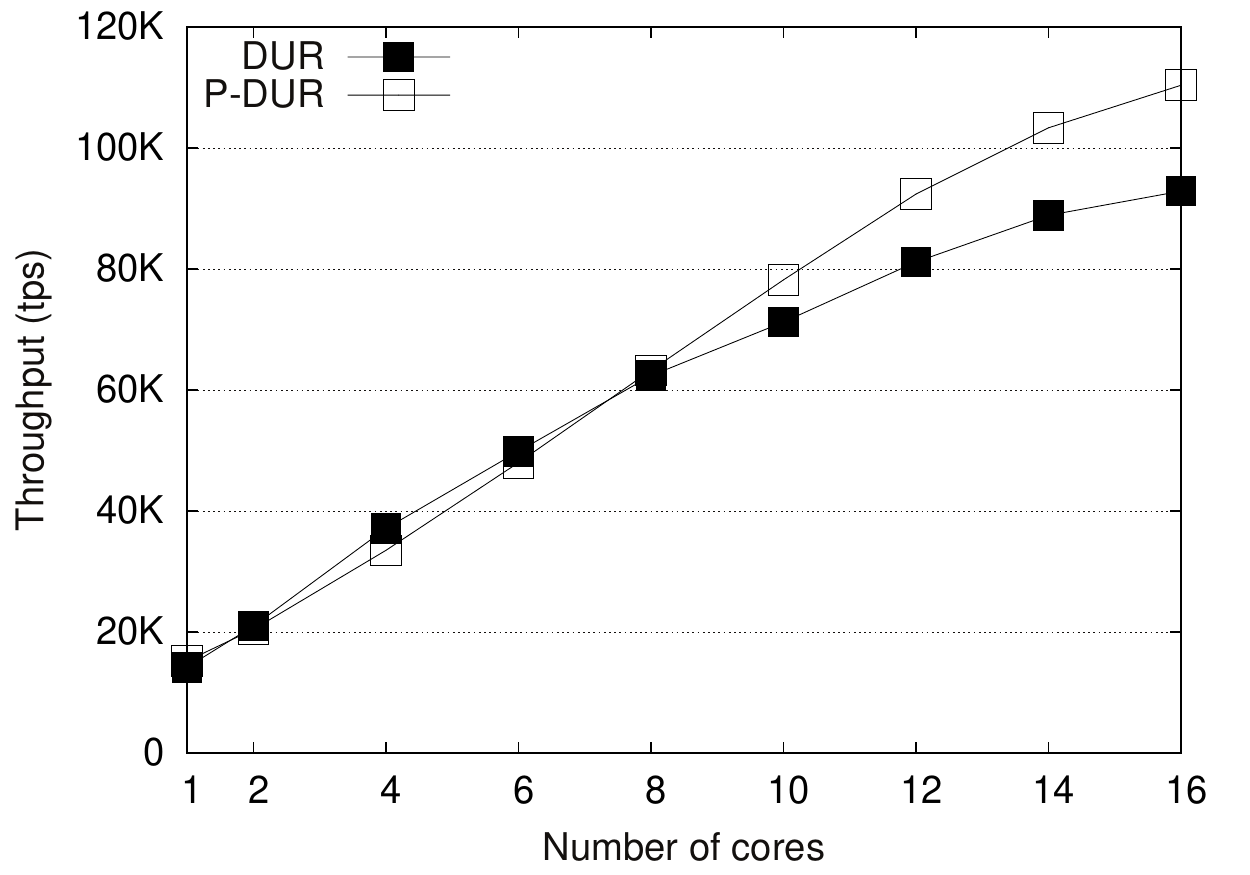} 
      \includegraphics[width=\sizefactor\columnwidth]{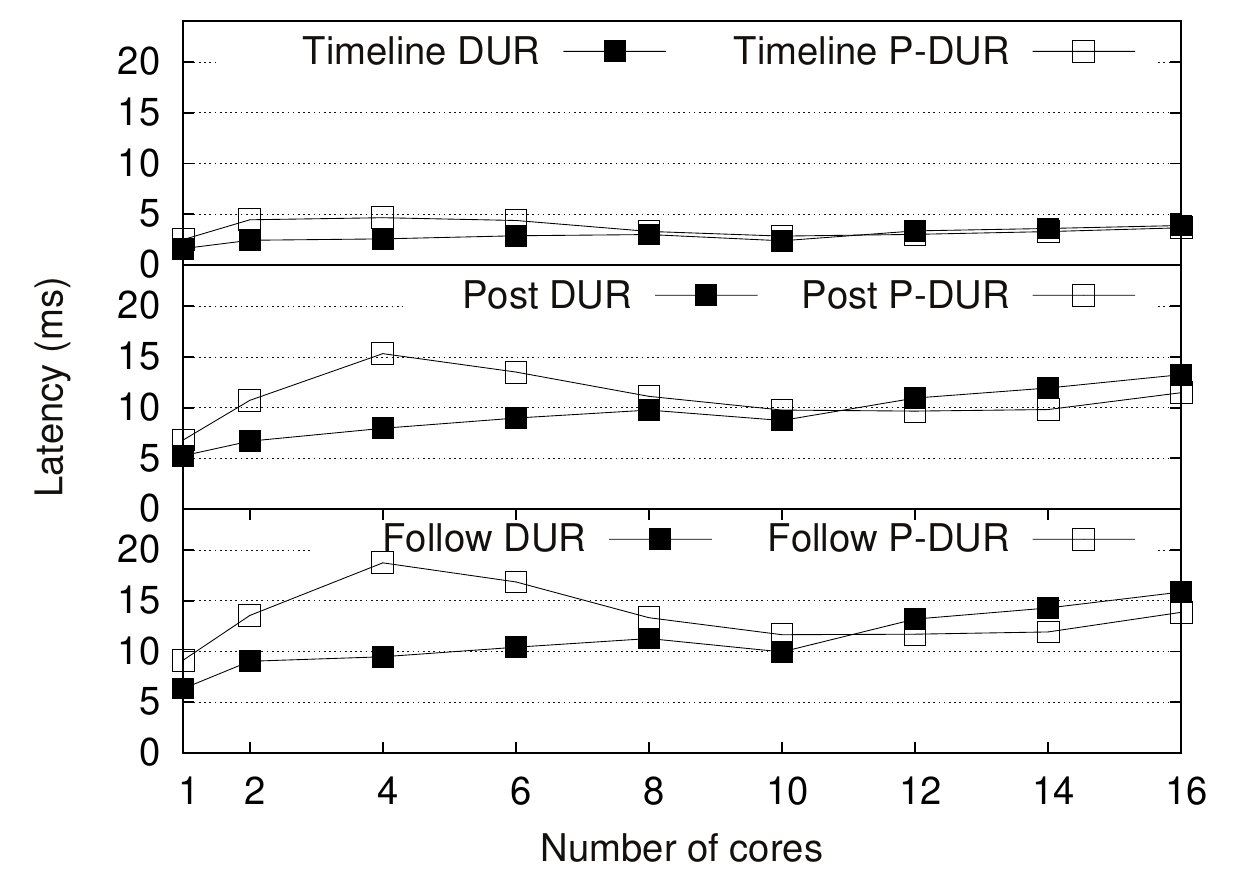} \\
    \end{tabular}
    \caption{\small Throughput and latency using the Twitter-like benchmark.}
    \label{twitter}
  \end{center}
	\vspace{-2mm}
\end{figure*}

%
We report results for a mix of 50\% timeline, 40\% post and 10\% follow transactions (see Figure~\ref{twitter}). 
Follow transactions are cross-partition with 50\% probability.
In these experiments, we populate the database with 420 thousand users. We generate random posts of length between 10 and 50 characters.

%
As discussed previously, DUR scales read-only transactions perfectly, as they are entirely local to a replica. 
Since the workload has 50\% of timeline operations (which are read-only), DUR scales as well as P-DUR up to 8 replicas. 
Past 8 replicas, update operations start affecting DUR's scalability. 
P-DUR scales well up to 14 cores with a slight decrease with 16 cores. 
Notice that in configurations with 16 cores, some cores are shared with operating systems tasks; 
we expect that in a node with more than 16 cores P-DUR would scale further.
Figure~\ref{twitter} also shows the latency for each type of operation. 
Timeline operations have a lower latency than the other two operation types, since timeline operations are read-only transactions and post and follow are update transactions.


\section{Related work}
\label{sec:related-work}

P-DUR is at the intersection of two areas of active research: multi-core databases and database replication.
In this section, we review work related to P-DUR in these two areas.

\subsection{Databases on multiple cores}

Traditional database management systems are known for not taking full advantage of modern multi-core hardware, despite the fact that database engines are multithreaded, typically to hide the latency of disk access so that multiple concurrent transactions can execute while other transactions are blocked waiting for I/O. 
Commodity hardware, equipped with large memories and fast CPUs, can nowadays easily keep OLTP workloads in memory and process a single transaction in the order of milliseconds. 
This has been observed in many experiments published in the literature \cite{Harizopoulos:2008, Johnson:2009, Salomie:2011, Stonebraker:2007}. 
Harizopoulos et al. \cite{Harizopoulos:2008} observed that in a typical database, features like buffer pool management, logging, locking, and latching introduce considerable overhead. 
Their experiments show that by removing those features, the resulting single-threaded database was able to perform OLTP transactions 20 times faster. Similarly, Stonebraker et al. \cite{Stonebraker:2007} showed that their protoype database, H-Store, could outperform a commercial database by almost two orders of magnitude on a TPC-C-like workload. 
These observations suggest that modern database architectures should take advantage of multi-core environments by treating a single multi-core machine as a shared-nothing cluster. 
This is the approach adopted by P-DUR.

Salomie et al. \cite{Salomie:2011} proposed Multimed, a middleware layer based on well-known distributed systems techniques to deploy a distributed database on a multi-core machine. 
Multimed uses lazy replication between a master node, which runs on a set of cores, and several satellite nodes running on different sets of cores. 
The master node accepts update transactions and lazily propagates updates to the satellites. 
This is similar to P-DUR, in which logically partitioned database instances are assigned to different cores. 
In contrast to Multimed, P-DUR does not rely on a single master, a potential bottleneck for update transactions. 
Jonhson et al. \cite{Johnson:2009} explore a different direction with Shore-MT. Shore-MT is an improved version of the Shore database \cite{Carey:1994}. Shore-MT employs optimized logging, locking and buffer management to remove major bottlenecks by careful synchronization of threads. 
The shared-everything approach of Shore-MT is pessimistic: synchronization has to be performed even if the workload has no contention at all, harming performance in the case of multi-socket multi-core CPUs \cite{Porobic:2012}.

\subsection{Database replication}

The literature on database replication is vast~\cite{CBPS10,KJP10}.
In the following, we focus on techniques based on the deferred update replication approach.
Many replication protocols for deferred update replication have been proposed in the past, most assuming a fully replicated database (e.g., \cite{AAES97, KA00a, LKPM+05, PMJP+05, Pedone:2003,ZP08}), which as we argued in Section~\ref{sec:deferred_update_replication} has inherent scalability limitations.

In order to improve performance, some protocols provide partial replication in the context of deferred update replication (e.g., \cite{SSP06, SPM+05, SPOM01}).
With partial replication, only the servers accessed by a transaction are involved in the execution and termination of the transaction, as opposed to all servers.
To commit a transaction, however, the protocols in \cite{SSP06}, \cite{SPM+05} and \cite{SPOM01} atomically broadcast all transactions to all participants. 
When delivering a transaction, a server simply discards those transactions that do not read or write items that are replicated locally.
Some protocols take full advantage of partial replication by relying on an atomic multicast primitive (e.g., \cite{FI01, SSP10}) or on a two-phase commit-like termination of transactions (e.g., \cite{SPJ12}).
Fritzke et al.~\cite{FI01} describe a protocol where each read operation of a transaction is multicast to all concerned partitions, and write operations are batched and multicast at commit time to the partitions concerned by the transaction.
P-Store~\cite{SSP10} implements deferred update replication with optimizations for wide-area networks. Upon commit, a transaction is multicast to the partitions containing items read or written by the transaction; partitions certify the transaction and exchange their votes.
Sciascia et al.~\cite{SPJ12} divide the database into partitions and each partition is fully replicated using deferred update replication. 
Similarly to P-Store, transactions that access multiple partitions are terminated using a two-phase commit-like protocol.
Differently from P-DUR, none of these protocols have been considered and evaluated on multi-core environments.


\section{Conclusion}
\label{sec:conclusion}

This paper revisited deferred update replication (DUR) in light of modern architecture trends, namely, multi-core servers.
Deferred update replication is a well-established replication technique, at the core of many database replication protocols.
While DUR has performance advantages when compared to other replication techniques, such as primary-backup and state-machine replication, it provides limited scalability with the number of replicas (i.e., poor scale out).
To the best of our knowledge, no previous work has proposed a multi-threaded solution to deferred update replication, that is, a solution that can scale up with multi-core architectures.
The real challenge lies in allowing concurrent termination of transactions, since certification for strong consistency implies some level of serialization.
Parallel Deferred Update Replication (P-DUR), the approach proposed in the paper, allows concurrency at both the execution and termination of transactions.
We have shown that P-DUR has very good performance in workloads dominated by single-partition transactions.


\bibliographystyle{ieeetr}
\bibliography{main}

\clearpage
\section*{Appendix}
\label{sec:appendix}

\label{sub:algorithm_correctness}

In this section, we argue that any execution of P-DUR, as seen by its clients, is serializable.
We first show that history $H$ with transactions committed at the servers contains all committed transactions submitted by the clients.
Let $t$ be a committed transaction, possibly cross-partition, that accesses partitions in $Q$, and $p$ a process in $s$ responsible for a partition in $Q$.
Since $t$ commits, $p$ received a commit vote from every other process $q$ in $s$ also responsible for $t$.
Therefore, all partitions involved in $t$ at correct servers have committed $t$.

We now show that for every execution $H$, there is a serial history $H_s$ with the same transactions that satisfies the following property:
if $t$ reads an item that was most recently updated by $t'$ in $H$ (i.e., ``$t$ reads from $t'$''), then $t$ reads the same item from $t'$ in $H_s$; in other words, $H$ and $H_s$ are \emph{equivalent}.
History $H$ is equivalent to a serial history $H_s$ where every two transactions $t$ and $t'$ in $H$ are disposed as follows in $H_s$: 
If $t$ and $t'$ \emph{access some common partition}, then $t$ precedes $t'$ in $H_s$ if and only if $t$ is delivered before $t'$; from the order property of atomic multicast, this is well-defined even if $t$ and $t'$ access multiple partitions in common. 
If $t$ and $t'$ \emph{do not access any common partition}, then $t$ precedes $t'$ in $H_s$ if and only if the unique identifier of $t$ is bigger than the unique identifier of $t'$.

Assume $t'$ reads some item $x$ from $t$ in $H$.
We must show that $t'$ reads $x$ from $t$ in $H_s$.
Since $H_s$ is serial, this can only happen if the two following conditions hold: (a)~$t$ precedes $t'$ in $H_s$ and (b)~no transaction $u$ that writes $x$ succeeds $t$ and precedes $t'$ in $H_s$.
Note that $t$ and $t'$ access at least one common partition, implemented by process $p$, the partition that contains $x$.

\emph{Condition (a).} 
Since transactions only read committed data in P-DUR, it follows that when $t'$ reads $x$, $t$ has already committed. In order to commit, $t$ must be first delivered. Thus, $t$ is delivered before $t'$, and $t$ precedes $t'$ in $H_s$.

\emph{Condition (b).} 
Assume for a contradiction that there exists a transaction $u$ that writes $x$, succeeds $t$, and precedes $t'$ in $H_s$.
Thus, $u$ is delivered after $t$ and before $t'$.
As a consequence, when $t'$ is certified, it is checked against $t$ and $u$.
Since $x \in t'.rs \cap u.ws$, it follows that $t'$ fails certification, a contradiction since $t'$ is a committed transaction.

\end{document}